\documentclass[lettersize,journal]{IEEEtran}
\usepackage{amsmath,amsfonts}
\usepackage{array}
\usepackage[caption=false,font=normalsize,labelfont=sf,textfont=sf]{subfig}
\usepackage{textcomp}
\usepackage{stfloats}
\usepackage{url}
\usepackage{verbatim}
\usepackage{graphicx}
\usepackage{cite}
\usepackage{soul}
\usepackage{balance}
\usepackage{algorithm}
\usepackage{algpseudocode}
\usepackage{booktabs}

\usepackage[most]{tcolorbox}
\tcbset{colback=blue!5!white, colframe=blue!75!black, arc=3mm, boxrule=0.8pt}

\usepackage[acronym,toc,shortcuts]{glossaries}

\makeglossaries
\newacronym{3GPP}{3GPP}{3rd Generation Partnership Project }
\newacronym{5G}{5G}{Fifth Generation}
\newacronym{6G}{6G}{Sixth Generation}
\newacronym{AAA}{AAA}{Authentication, Authorization and Accounting}
\newacronym{AGV}{AGV}{Automated Guided Vehicle}
\newacronym{AI}{AI}{Artificial Intelligence}
\newacronym{AP}{AP}{Access Point}
\newacronym{API}{API}{Application Programming Interface}
\newacronym{AMF}{AMF}{Access and Mobility Management Function}
\newacronym{APN}{APN}{Access Point Name}
\newacronym{AR}{AR}{Augmented Reality}
\newacronym{ANR}{ANR}{Automatic Neighbour Relationship}
\newacronym{ASN}{ASN}{Abstract Syntax Notation}
\newacronym{AWS}{AWS}{Amazon Web Services}
\newacronym{BaaS}{BaaS}{Backend-as-a-Service}
\newacronym{BS}{BS}{Base Station}
\newacronym{BB}{BB}{BaseBand}
\newacronym{BBU}{BBU}{BaseBand Unit}
\newacronym{BCN}{BCN}{Blockchain Network}
\newacronym{BER}{BER}{Bit Error Rate}
\newacronym{BSSID}{BSSID}{Basic Service Set Identification}
\newacronym{CaaS}{CaaS}{Containers as a Service}
\newacronym{CaPEX}{CaPEX}{Capital Expenditure}
\newacronym{ClaaS}{ClaaS}{Cloud as a Service}
\newacronym{CA}{CA}{Carrier Aggregation}
\newacronym{CAT}{CAT}{Capacity-Aware TOPSIS}
\newacronym{CAPEX}{CAPEX}{Capital Expenditure}
\newacronym{CD}{CD}{Continuous Development}
\newacronym{CDN}{CDN}{Content Delivery Network}
\newacronym{CEP}{CEP}{Complex Event Processing}
\newacronym{CELL-ID}{CELL-ID}{cell identification ID}
\newacronym{CGI}{CGI}{Cell Global Identification}
\newacronym{CI}{CI}{Continuous Integration}
\newacronym{CLSM}{CLSM}{Closed loop spatial multiplexing}
\newacronym{CQI}{CQI}{Channel Quality Indicator}
\newacronym{CN}{CN}{Core Network}
\newacronym{CNF}{CNF}{Cloud-native Network Function}
\newacronym{CoMP}{CoMP}{Coordinated Multipoint Transmission/Reception}
\newacronym{CP}{CP}{Connection Point}
\newacronym{CPS}{CPS}{Cyber-Physical System}
\newacronym{CPU}{CPU}{Central Processing Unit}
\newacronym{CNA}{CNA}{Cloud-Native Application}
\newacronym{CNFC}{CNFC}{CNF Component}
\newacronym{CNCF}{CNCF}{Cloud Native Computing Foundation}
\newacronym{CNFVI}{CNFVI}{Cloud-native NFV}
\newacronym{CNN}{CNN}{Convolutional Neural Networks}
\newacronym{CNTT}{CNTT}{Cloud INfrastructure Telco Taskforce}
\newacronym{CL}{CL}{Closed-Loop}
\newacronym{CSMF}{CSMF}{Core Slice Management Function}
\newacronym{CU}{CU}{Central Unit}
\newacronym{COTS}{COTS}{Commercial off-the-shelf}
\newacronym{CPRI}{CPRI}{Common Public Radio Interface}
\newacronym{CRM}{CRM}{Customer Relationship Management}
\newacronym{CS}{CS}{central scheduler}
\newacronym{CSI}{CSI}{Channel Status Information}
\newacronym{CSP}{CSP}{Communication Service Provider}
\newacronym{eNB}{eNB}{evolved Node-B}
\newacronym{D2D}{D2D}{Device-to-Device}
\newacronym{DC}{DC}{Data Center}
\newacronym{DCN}{DCN}{Dedicated Core Network}
\newacronym{DCS}{DCS}{Distributed Control System}
\newacronym{DECOR}{DECOR}{Dedicated Core Network}
\newacronym{DevOps}{DevOps}{Development and Operations}
\newacronym{DTLS}{DTLS}{Datagram Transport Layer Security}
\newacronym{DL}{DL}{Downlink}
\newacronym{DLT}{DLT}{Distributed Ledger Technology}
\newacronym{DID}{DID}{Decentralized Identifier}
\newacronym{DMO}{DMO}{Domain Management Orchestrator}
\newacronym{DNS}{DNS}{Domain Name Server}
\newacronym{DRL}{DRL}{Deep Reinforcement Learning}
\newacronym{DT}{DT}{Digital Twin}
\newacronym{DU}{DU}{Distributed Unit}
\newacronym{DMM}{DMM}{Distributed Mobility Management}
\newacronym{DPU}{DPU}{Data Processing Unit}
\newacronym{DPDK}{DPDK}{Data Plane Development Kit}
\newacronym{eBPF}{eBPF}{extended Berkeley Packet Filter}
\newacronym{eCPRI}{eCPRI}{enhanced CPRI}
\newacronym{ECA}{ECA}{Event-Condition-Action}
\newacronym{eMBB}{eMBB}{enhanced Mobile Broadband}
\newacronym{EMI}{EMI}{Electromagnetic Interference}
\newacronym{EMC}{EMC}{Electromagnetic Compatibility}
\newacronym{eNodeB}{eNodeB}{evolved Node-B}
\newacronym{eSIM}{eSIM}{Electronic SIM}
\newacronym{E2E}{E2E}{End-to-End}
\newacronym{EPC}{EPC}{Evolved Packet Core}
\newacronym{EPS}{EPS}{Evolved Packet System}
\newacronym{E-RAB}{E-RAB}{E-UTRAN Radio Access Bearer}
\newacronym{ERP}{ERP}{Enterprise Resource Planning}
\newacronym{ETSI}{ETSI}{European Telecommunications Standards Institute}
\newacronym{FaaS}{FaaS}{Functions as a Service}
\newacronym{FCC}{FCC}{Federal Communications Commission}
\newacronym{FDD}{FDD}{Frequency Division Duplexing}
\newacronym{FEM}{FEM}{Flow Extraction Manager}
\newacronym{FoF}{FoF}{Factories-of-the-Future}
\newacronym{FMC}{FMC}{Fixed Mobile Convergence}
\newacronym{FPGA}{FPGA}{Field-Programmable Gate Array}
\newacronym{FTP}{FTP}{File Transfer Protocol}
\newacronym{FCAPS}{FCAPS}{Fault, Configuration, Accounting, Performance and Security}
\newacronym{GCP}{GCP}{Google Cloud Platform}
\newacronym{GDPR}{GDPR}{General Data Protection Regulation}
\newacronym{GGSN}{GGSN}{Gateway GPRS Support Node}
\newacronym{GPRS}{GPRS}{General packet radio service}
\newacronym{GPS}{GPS}{Global Positioning System}
\newacronym{gRPC}{gRPC}{gRPC Remote Procedure Calls}
\newacronym{GTP}{GTP}{GPRS Tunneling Protocol}
\newacronym{HetNet}{HetNet}{heterogeneous network}
\newacronym{HO}{HO}{Handover}
\newacronym{HRoT}{HRoT}{Hardware Root of Trust}
\newacronym{HSS}{HSS}{Home Subscriber Station}
\newacronym{HTTP}{HTTP}{Hypertext Transfer Protocol}
\newacronym{HD}{HD}{High-Definition}
\newacronym{HDFS}{HDFS}{Hadoop Distributed File System}
\newacronym{HiveQL}{HiveQL}{Hive Query language}
\newacronym{HMI}{HMI}{Human-Machine Interface}
\newacronym{HNSM}{HNSM}{Host-based Network Security Manager}
\newacronym{HSPA}{HSPA}{High Speed Packet Access}
\newacronym{HSM}{HSM}{Hardware Security Module}
\newacronym{IAB}{IAB}{Integrated Access and Backhaul}
\newacronym{IAM}{IAM}{Identity and Access Management}
\newacronym{IaC}{IaC}{Infrastructure as Code}
\newacronym{IaaS}{IaaS}{Infrastructure as a Service}
\newacronym{IBLER}{IBLER}{Initial Block Error Rate}
\newacronym{ICIC}{ICIC}{inter-cell interference coordination}
\newacronym{ICN}{ICN}{information-centric network}
\newacronym{ICT}{ICT}{Information and Communication Technologies}
\newacronym{IDM}{IDM}{Infrastructure Domain Manager}
\newacronym{IDMO}{IDMO}{Inter-Domain Management Orchestrator}
\newacronym{IEC}{IEC}{International Electrotechnical Commission}
\newacronym{IEEE}{IEEE}{Institute of Electrical and Electronics Engineers}
\newacronym{IETF}{IETF}{Internet Engineering Task Force}
\newacronym{IGA}{IGA}{Identity Governance and Administration}
\newacronym{IIoT}{IIoT}{Industrial IoT}
\newacronym{IMSI}{IMSI}{International Mobile Subscriber Identity}
\newacronym{IMEI}{IMEI}{International Mobile Station Equipment Identity}
\newacronym{IMS}{IMS}{IP Multimedia Subsystem}
\newacronym{IDoT}{IDoT}{Identity of Things}
\newacronym{ID}{ID}{Identifier}
\newacronym{I/O}{I/O}{Input/Output}
\newacronym{ICMP}{ICMP}{Internet Control Message Protocol}
\newacronym{ISA}{ISA}{International Society of Automation}
\newacronym{IoT}{IoT}{Internet of Things}
\newacronym{IP}{IP}{Internet Protocol}
\newacronym{IRS}{IRS}{Intelligent Reflective Surface}
\newacronym{ISO}{ISO}{International Organization for Standardization}
\newacronym{ITU}{ITU}{International Telecommunication Union}
\newacronym{IT}{IT}{Information Technology}
\newacronym{ITS}{ITS}{Intelligent Transportation Systems}
\newacronym{Li-Fi}{Li-Fi}{Light Fidelity}
\newacronym{L2}{L2}{Layer 2}
\newacronym{LBO}{LBO}{Local Break Out}
\newacronym{LCM}{LCM}{Life Cycle Management}
\newacronym{LDPC}{LDPC}{Low Density Parity Check}
\newacronym{LPWAN}{LPWAN}{Low Power Wide Area Network}
\newacronym{L3}{L3}{Layer 3}
\newacronym{GBR}{GBR}{Guaranteed Bit Rate}
\newacronym{GLUE}{GLUE}{General Language Understanding Evaluation}
\newacronym{MI}{MI}{Middleware Interface}
\newacronym{JSON}{JSON}{JavaScript Object Notation}
\newacronym{IMT}{IMT}{International Mobile Telecommunications}
\newacronym{K8s}{K8s}{Kubernetes}
\newacronym{KMS}{KMS}{Key Management Service}
\newacronym{KPI}{KPI}{Key Performance Indicator}
\newacronym{KYC}{KYC}{Know Your Customer}
\newacronym{LA}{LA}{Location Area}
\newacronym{LAC}{LAC}{location area code}
\newacronym{LDAP}{LDAP}{Lightweight Directory Access Protocol}
\newacronym{LMA}{LMA}{Local Mobility Anchor}
\newacronym{LTE}{LTE}{long term evolution}
\newacronym{mMTC}{mMTC}{massive Machine-Type Communications}
\newacronym{MADM}{MADM}{Multiple Attribute Decision Making}
\newacronym{MCC}{MCC}{Mobile Country Code}
\newacronym{MCS}{MCS}{Modulation Coding Scheme}
\newacronym{MNC}{MNC}{Mobile Network Code}
\newacronym{MIMO}{MIMO}{multiple-input multiple-output}
\newacronym{MAG}{MAG}{Mobile Access Gateway}
\newacronym{MAAR}{MAAR}{Mobility Anchor and Access Router}
\newacronym{MANO}{MANO}{Management and Orchestration}
\newacronym{MEC}{MEC}{Multi-access edge computing}
\newacronym{MFA}{MFA}{Multi-Factor Authentication}
\newacronym{ML}{ML}{Machine Learning}
\newacronym{MME}{MME}{Mobility Management Entity}
\newacronym{MDM}{MDM}{Mobile Device Management}
\newacronym{MN}{MN}{Mobile Node}
\newacronym{MTC}{MTC}{machine Type Communication}
\newacronym{MNO}{MNO}{Mobile Network Operator}
\newacronym{MR}{MR}{Mixed Reality}
\newacronym{MSISDN}{MSISDN}{Mobile Station International Subscriber Directory Number}
\newacronym{NAT}{NAT}{Network Address Translation}
\newacronym{NBI}{NBI}{NorthBound Interface}
\newacronym{NC}{NC}{Network Coding}
\newacronym{NDT}{NDT}{Network Digital Twin}
\newacronym{NEF}{NEF}{Network Exposure Function}
\newacronym{NF}{NF}{Network Function}
\newacronym{NFV}{NFV}{Network Functions Virtualization}
\newacronym{NFVO}{NFVO}{Network Functions Virtualization Orchestrator}
\newacronym{NFVI}{NFVI}{Network Functions Virtualization Infrastructure}
\newacronym{NIST}{NIST}{National Institute of Standards and Technology}
\newacronym{NIC}{NIC}{Network Interface Card}
\newacronym{NLP}{NLP}{Natural Language Processing}
\newacronym{NLU}{NLU}{Natural Language Understanding}
\newacronym{NOMA}{NOMA}{Non-Orthogonal Multiple Access}
\newacronym{NoSQL}{NoSQL}{Not Only SQL}
\newacronym{NPN}{NPN}{Non-Public Network}
\newacronym{NR}{NR}{New Radio}
\newacronym{NRF}{NRF}{Network Repository Function}
\newacronym{NS}{NS}{Network Service}
\newacronym{NSSMF}{NSSMF}{Network Slice Subnet Management Function}
\newacronym{NSMF}{NSMF}{Network Slice Selection Function}
\newacronym{NTN}{NTN}{Non-Terrestrial Networks}
\newacronym{QoS}{QoS}{Quality-of-Service}
\newacronym{QoE}{QoE}{Quality-of-Experience}
\newacronym{PaaS}{PaaS}{Platform as a Service}
\newacronym{PNF}{PNF}{Physical Network Function}
\newacronym{PCB}{PCB}{Printed Circuit Board}
\newacronym{PDN}{PDN}{packet data network}
\newacronym{PI}{PI}{Provision Interface}
\newacronym{PID}{PID}{Proportional Integral Derivative}
\newacronym{PF}{PF}{Proportional Fair}
\newacronym{P-GW}{P-GW}{packet gateway}
\newacronym{PDP}{PDP}{Packet Data Protocol}
\newacronym{PDU}{PDU}{Packet Data Unit}
\newacronym{PFH}{PFH}{Probability of Failure on Demand per Hour}
\newacronym{PHY}{PHY}{physical layer}
\newacronym{PKI}{PKI}{Public Key Infrastructure}
\newacronym{PLC}{PLC}{Programmable Logic Controller}
\newacronym{PLMN}{PLMN}{Public Land Mobile Network}
\newacronym{PMIPv6}{PMIPv6}{Proxy Mobile IPv6}
\newacronym{PMI}{PMI}{Precoding Matrix Index}
\newacronym{PoP}{PoP}{Point of Presence}
\newacronym{PQC}{PQC}{Post Quantum Cryptography}
\newacronym{PRB}{PRB}{Physical Resource Block}
\newacronym{PUSCH}{PUSCH}{Physical Uplink Shared Channel}
\newacronym{QAM}{QAM}{Quadrature amplitude modulation}
\newacronym{QCI}{QCI}{QoS Class Identifier}
\newacronym{QKD}{QKD}{Quantum Key Distribution}
\newacronym{RA}{RA}{Routing Area}
\newacronym{RAMI}{RAMI}{Reference Architectural Model Industrie}
\newacronym{RB}{RB}{Resource Block}
\newacronym{RES}{RES}{Renewable Energy Sources}
\newacronym{REST}{REST}{Representational State Transfer}
\newacronym{RPC}{RPC}{Remote Procedure Call}
\newacronym{RI}{RI}{Rank Indicator}
\newacronym{RAN}{RAN}{Radio Access Network}
\newacronym{RBAC}{RBAC}{Role-Based Access Control}
\newacronym{RFC}{RFC}{Request for Comment}
\newacronym{RF}{RF}{Radio Frequency}
\newacronym{RFID}{RFID}{Radio-frequency identification}
\newacronym{RRC}{RRC}{Radio Resource Control}
\newacronym{RRU}{RRU}{Remote Radio Unit}
\newacronym{RNC}{RNC}{radio network controller}
\newacronym{RNN}{RNN}{Recurrent Neural Networks}
\newacronym{RSSI}{RSSI}{Received Signal Strength Indicator}
\newacronym{RSRP}{RSRP}{Reference Signal Received Power}
\newacronym{RTT}{RTT}{Round Trip Time}
\newacronym{OSI}{OSI}{Open Systems Interconnection}
\newacronym{OPEX}{OPEX}{Operational Expenditure}
\newacronym{OAM}{OAM}{Operation, Administration and Management}
\newacronym{OL}{OL}{Open-Loop}
\newacronym{ONAP}{ONAP}{Open Networking Automation Platform}
\newacronym{ONF}{ONF}{Open Networking Foundation}
\newacronym{ONOS}{ONOS}{Open Network Operating System}
\newacronym{O-RAN}{O-RAN}{Open Radio Access Network}
\newacronym{OpEX}{OpPEX}{Operational Expenditure}
\newacronym{OPNFV}{OPNFV}{Open Platform for Network Functions Virtualization}
\newacronym{OS}{OS}{Operating System}
\newacronym{OSM}{OSM}{Open Source MANO}
\newacronym{OSS}{OSS}{Operational Support Systems}
\newacronym{OT}{OT}{Operational Technology}
\newacronym{OVP}{OVP}{OPNFV Verification Program}
\newacronym{OTT}{OTT}{over-the-top}
\newacronym{OTP}{OTP}{one-time-password}
\newacronym{QR}{QR}{Quick Response}
\newacronym{PET}{PET}{Privacy-Enhancing Technology}
\newacronym{RoT}{RoT}{Root of Trust}
\newacronym{SaaS}{SaaS}{Software as a Service}
\newacronym{SA}{SA}{Stand Alone}
\newacronym{SAML}{SAML}{Security Assertion Markup Language}
\newacronym{SB}{SB}{Service Bus}
\newacronym{SI}{SI}{Service Interface}
\newacronym{SIL}{SIL}{Safety Integrity Level}
\newacronym{SAC}{SAC}{service area code}
\newacronym{SBA}{SBA}{Service Based Architecture}
\newacronym{SBI}{SBI}{Service Based Interface}
\newacronym{SBMA}{SBMA}{Service-Based Management Architecture}
\newacronym{SD}{SD}{service Discovery}
\newacronym{SCMA}{SCMA}{Sparse Code Multiple Access}
\newacronym{SDLC}{SDLC}{Software Development Life Cycle}
\newacronym{SFC}{SFC}{Service Function Chaining}
\newacronym{SIM}{SIM}{Subscriber Identity Module}
\newacronym{SMI}{SMI}{Service Management Interface}
\newacronym{SLA}{SLA}{Service Level Agreement}
\newacronym{SDN}{SDN}{Software Defined Networking}
\newacronym{SDK}{SDK}{Software Development Kit}
\newacronym{SDO}{SDO}{Standards Developing Organization}
\newacronym{SOA}{SOA}{Service Oriented Architecture}
\newacronym{SFN}{SFN}{Single Frequency Network}
\newacronym{SEI}{SEI}{Service External Interface}
\newacronym{SMW}{SMW}{Software Middleware}
\newacronym{S-GW}{S-GW}{serving gateway}
\newacronym{SR-IOV}{SR-IOV}{Single Root Input Output Virtualization}
\newacronym{SINR}{SINR}{signal-to-interference-plus-noise ratio}
\newacronym{SGSN}{SGSN}{Serving GPRS Support Node}
\newacronym{SSI}{SSI}{Self Sovereign Identity}
\newacronym{SSID}{SSID}{Service Set Identification}
\newacronym{SSL}{SSL}{Secure Socket Layer}
\newacronym{STO}{STO}{Safe Torque Off}
\newacronym{SVD}{SVD}{singular value decomposition}
\newacronym{SW}{SW}{Software}
\newacronym{SU}{SU}{Service Unit}
\newacronym{TAR}{TAR}{Tape Archive}
\newacronym{TCP}{TCP}{transport control protocol}
\newacronym{TLS}{TLS}{Transport Layer Security}
\newacronym{TCO}{TCO}{total cost ownership}
\newacronym{TDD}{TDD}{Time Division Duplexing}
\newacronym{TM}{TM}{transmission mode}
\newacronym{TPM}{TPM}{Trusted Platform Module}
\newacronym{TSN}{TSN}{Time Sensitive Network}
\newacronym{TUG}{TUG}{Telecom User Group}
\newacronym{TEID}{TEID}{tunnel endpoint identifier}
\newacronym{UAV}{UAV}{Unmanned Aerial Vehicle}
\newacronym{UDN}{UDN}{Ultra Dense Network}
\newacronym{UMTS}{UMTS}{Universal Mobile Telecommunications Service} 
\newacronym{UE}{UE}{user equipment}
\newacronym{UL}{UL}{Uplink}
\newacronym{UP}{UP}{User Plane}
\newacronym{URL}{URL}{Uniform Resource Locator}
\newacronym{uRLLC}{uRLLC}{ultra-Reliable Low-Latency Communication}
\newacronym{UDP}{UDP}{User Datagram Protocol}
\newacronym{UPF}{UPF}{User Plane Function}
\newacronym{VC}{VC}{Verifiable Credential}
\newacronym{VIM}{VIM}{Virtual Infrastructure Manager}
\newacronym{VL}{VL}{Virtual Link}
\newacronym{VF}{VF}{Virtual Function}
\newacronym{VNF}{VNF}{Virtual Network Function}
\newacronym{VPN}{VPN}{Virtual Private Network}
\newacronym{vRAN}{vRAN}{Virtual Radio Access Network}
\newacronym{VNFC}{VNFC}{Virtual Network Function Component}
\newacronym{VNFM}{VNFM}{Virtual Network Function Manager}
\newacronym{VNFFG}{VNFFG}{Virtualized Network Function Forwarding Graphs}
\newacronym{VM}{VM}{Virtual Machine}
\newacronym{vRC}{vRC}{Virtual Radio Controller}
\newacronym{vPP}{vPP}{Virtual Packet Processor}
\newacronym{VR}{VR}{Virtual Reality}
\newacronym{XR}{XR}{Extended Reality}
\newacronym{WiFi}{WiFi}{Wireless Fidelity}
\newacronym{WLAN}{WLAN}{Wireless Local Area Network}
\newacronym{ZSM}{ZSM}{Zero touch network \& Service Management}
\newacronym{ZKP}{ZKP}{Zero-Knowledge Proof}
\newacronym{ZTA}{ZTA}{Zero-Trust Architecture}

\begin{document}

\title{Modular Multi-Domain Digital Twin Architecture: Sustainable Intent-Driven 6G Management}

\author{Berk Buzcu,~\IEEEmembership{Student Member,~IEEE,}, 
        Marcin Pakula, 
        Gevher Yesevi Keskin, 
        Laura Finarelli,~\IEEEmembership{Student Member,~IEEE,} 
        Gianluca Rizzo,~\IEEEmembership{Senior Member,~IEEE,} 
        Engin Zeydan,~\IEEEmembership{Senior Member,~IEEE,} 
        Jorge Baranda,~\IEEEmembership{Senior Member,~IEEE,} 
        Aitor Alcázar-Fernández, 
        Javier Velázquez-Martínez, 
        Luis M. Contreras, 
        Gil Kedar, 
        Efi Dvir, 
        Paweł Kryszkiewicz,~\IEEEmembership{Senior Member,~IEEE} 

\thanks{This work is supported by UNITY-6G project, co-funded from European Union’s Horizon Europe Smart Networks and Services
Joint Undertaking (SNS JU) research and innovation programme
under the Grant Agreement No 101192650 and from the Swiss State Secretariat for Education, Research and Innovation (SERI).}        
\thanks{B. Buzcu and L. Finarelli are with the University of Applied Sciences and Arts Western Switzerland (HES-SO Valais/Wallis), Switzerland, and with Technische Universität Berlin, Germany. (e-mail: berk.buzcu@hevs.ch, laura.finarelli@hevs.ch)\\
M. Pakula and P. Kryszkiewicz are with the Poznań University of Technology (PUT), Poland (e-mail: marcin.pakula@doctorate.put.poznan.pl, pawel.kryszkiewicz@put.poznan.pl). \\
G. Yesevi Keskin is with Centrum Wiskunde \& Informatica (CWI), Amsterdam, The Netherlands, and with Eindhoven University of Technology, The Netherlands. (e-mail: g.yesevi.keskin@tue.nl)\\
E. Zeydan and J. Baranda are with the Centre Tecnològic de Telecomunicacions de Catalunya (CTTC), Castelldefels, Spain. (e-mail: ezeydan@cttc.es, jorge.baranda@cttc.es) \\
A. Alcázar-Fernández, J. Velázquez-Martínez, and L. M. Contreras are with Telefónica Innovación Digital, Madrid, Spain. (e-mail: aitor.alcazarfernandez@telefonica.com, j.velamar@telefonica.com, luismiguel.contrerasmurillo@telefonica.com)\\
G. Kedar is with Ceragon Networks, Israel (e-mail: gilke@ceragon.com). \\
G. Rizzo is with the University of Applied Sciences and Arts Western Switzerland (HES-SO Valais/Wallis), Switzerland, and with University of Turin, Italy. (e-mail: gianluca.rizzo@hevs.ch)}        
}

\markboth{Journal of \LaTeX\ Class Files,~Vol.~14, No.~8, August~2026}%
{Author \MakeLowercase{\textit{et al.}}: Digital Twin-Enabled 6G Wireless Networks}

\maketitle

\begin{abstract}
Future 6G networks will operate across distributed and heterogeneous domain infrastructures, making conventional single-domain management insufficient for proactive, trustworthy automation. Network Digital Twins (NDTs) enable what-if analysis, AI-assisted optimization, and risk-free validation of control actions before deployment, yet monolithic end-to-end twins remain impractical due to scalability, fidelity, and cross-domain coordination challenges. Accordingly, this paper proposes a Digital Twin-enabled 6G architecture that exposes NDT capabilities as a specialized service domain within a multi-domain orchestration framework built on a state-of-the-art service-based 6G architecture. A DT Orchestrator interprets \textit{predictive} and \textit{prescriptive} what-if queries and composes domain-specific DT modules and simulators on demand, while decision authority remains with the requesting entity. Furthermore, a generalized workflow covers telemetry synchronization, simulation-based decision support, and closed-loop execution. The framework is demonstrated through a green-networking use case that couples a system-level O-RAN cellular digital twin component with a two-stage solar-allocation simulator, evaluated over a 105-base-station deployment in Poznan using simulative datasets. Joint coverage and renewable optimization reduces daily grid consumption by 28.5\% with 32 solar panels at the diminishing-returns threshold, with 17 base stations identified as both coverage-active and high-priority solar candidates as evidence that cross-domain NDT coordination enables sustainable, intent-driven 6G network management.

\end{abstract}

\begin{IEEEkeywords}
6G, Digital Twin, Edge Computing, Network Virtualization, Artificial Intelligence.
\end{IEEEkeywords}


\section{Introduction}
The evolution toward new generations of cellular networks, and particularly \ac{6G}, requires a shift from reactive troubleshooting to proactive, AI-driven network management. Future networks will operate in highly dynamic and heterogeneous environments, integrating radio access, transport, core, and edge-cloud infrastructures while meeting stringent requirements for latency, reliability, scalability, and sustainability. Traditional management frameworks, constrained by isolated domains, \textit{telco-only} perspectives, static control policies, and offline optimization, are no longer sufficient.

To manage this complexity, \ac{6G} networks are expected to rely on the \ac{NDT} as a synchronized virtual representation of network assets, services, and operational states. The \ac{NDT} is not merely a monitoring tool, but a high-fidelity virtual environment that enables operators to simulate \textit{what-if} scenarios, assess alternative configurations, and validate AI-driven decisions before live deployment. In this role, the \ac{NDT} provides a risk-aware decision-support layer between network telemetry and operational control, improving the reliability of closed-loop automation. However, constructing a monolithic end-to-end \ac{NDT} for future 6G ecosystems remains impractical due to the scale, heterogeneity, and computational complexity of multi-domain environments. A more feasible approach is to coordinate multiple domain-specific twins, simulators, and analytical models under a common orchestration framework. In this work, we adopt the inter-domain management and orchestration architecture proposed in~\cite{BuzcuEtAlWONS2026}, which enables coordinated service delivery across heterogeneous administrative and technological domains.

Beyond conventional telecom optimization, \ac{NDT}s also create an opportunity to incorporate non-telecommunication information sources into network control loops. This is particularly relevant for sustainability-oriented 6G systems, where renewable energy availability, carbon-intensity indicators, and localized energy forecasts can influence network decisions. In this context, the \ac{NDT} enables networks to evolve from being merely energy-efficient, i.e., consuming less power, to becoming carbon-aware, i.e., consuming energy when and where it is most sustainable. The need for this capability is reinforced by the evolution of the energy sector. Modern power grids are no longer passive distribution infrastructures; they increasingly rely on forecasting, demand-response mechanisms, and autonomous management of decentralized \ac{RES}~\cite{AlEmari2025}. Consequently, the 6G network should not treat the grid as an opaque energy source, but as an intelligent external domain whose state can be exploited for network optimization. The \ac{NDT} acts as the mediator in this relationship, ingesting energy-domain information such as predicted carbon intensity or renewable energy surpluses and translating it into actionable network-management decisions.

Motivated by these challenges, this paper proposes a \ac{DT}-enabled multi-domain 6G architecture in which the \ac{NDT} acts as a specialized orchestration capability for scenario analysis, intent-driven optimization, and coordinated decision support. The framework integrates \ac{NDT} functionalities within a multi-domain orchestration layer, enabling coordinated management of heterogeneous network segments through structured what-if analysis and closed-loop control. Its applicability is demonstrated through a sustainability-oriented use case that combines network simulation outputs with renewable energy information to optimize base-station operation under energy-aware objectives.

The paper is organized as follows. Section~\ref{sec:related-work} reviews the state of the art on \glspl{DT} and multi-domain \ac{NDT} architectures, identifying the gap addressed in this work. Section~\ref{sec:background} presents the proposed multi-domain 6G orchestration architecture and describes how \ac{NDT} capabilities are integrated to support coordinated service management. Section~\ref{sec:proposed-approach} details the corresponding service workflow, including DT instantiation, what-if analysis, KPI-driven optimization, and closed-loop execution. Section~\ref{sec:green-network-uc} introduces the sustainability use case, where the \ac{NDT} Manager orchestrates the interaction between a system-level cellular RAN DT Module, and a energy simulator to resolve a cross-domain what-if scenario. Section~\ref{sec:evaluation} elaborates on the evaluation method of the framework. Finally, Section~\ref{sec:conclusions} concludes the paper.

\section{Related Work}
\label{sec:related-work}
This section reviews the evolution from general \ac{DT} concepts toward \ac{NDT} architectures for future wireless systems. First, it consolidates the foundational understanding of DTs as synchronized, purpose-driven virtual representations. Then, it analyses the role of NDTs in 6G networks, with emphasis on decision support, pre-validation, and trustworthy automation. Finally, it discusses the transition from scoped, domain-specific NDTs toward multi-domain NDT coordination, which directly motivates the architecture proposed in this work.

\subsection{Foundations of Digital Twin Architectures}

The concept of a DT originated as a virtual counterpart to a physical asset or system, continuously updated with operational data to support monitoring, analysis, prediction, and optimization. Over time, the literature has shifted from broad and sometimes inconsistent interpretations to a more operational understanding: a DT is not a static model, simulator, or visualization tool, but a synchronized and purpose-driven virtual representation that supports system-level reasoning. Its value lies not in reproducing every internal building block of the physical system, but in providing purpose-specific fidelity, i.e., accurately capturing the behaviours, KPIs, or decision outcomes relevant to the intended use case. 

For networking, this distinction is critical. Communication systems are distributed, dynamic, and multi-layered, making permanent full-fidelity replication impractical. Accordingly, recent NDT-oriented work increasingly frames the twin as a modular architectural capability composed of data collection and synchronization mechanisms, virtual models, analytics functions, lifecycle management procedures, and interfaces exposed to external consumers~\cite{khan_digital_2022, poorzare_network_2025, qin_machine_2024}. This interpretation is increasingly reflected in standardization efforts such as ITU-T Y.3090~\cite{ITUY3090}, the IRTF NMRG reference architecture draft~\cite{zhou_network_2025}, and 3GPP TS 28.561~\cite{3GPP_TS28561}, which move the discussion from abstract definitions toward scenario-aware, lifecycle-managed, and automation-oriented NDT realizations.

From this perspective, an NDT analysis evaluates the operational consequences of candidate network decisions before they are applied to the production system. Such analyses may address planning, configuration validation, fault assessment, congestion mitigation, recovery strategies, energy optimization, or ML-enabled control. The common objective is to reduce operational risk and improve the reliability of automated decision-making by evaluating the expected impact of alternative actions under realistic assumptions.

\subsection{Network Digital Twins in Future Wireless Networks (6G)}

In 6G, the NDT evolves from a supporting management tool into a core enabler of trustworthy, proactive, and automation-ready network operations. Future wireless systems must support dense multi-RAT deployments, cloud-edge-native service execution, AI-native control loops, service-aware optimization, and stringent requirements for latency, reliability, and energy efficiency. In these conditions, validating decisions directly on production infrastructure becomes increasingly costly, risky, and often impractical. The NDT addresses this limitation by providing a synchronized and scenario-aware environment where network behavior, service performance, and control actions can be evaluated before implementation in the real system~\cite{lin_6g_2023,campoy_digital_2025,tran_network_2025}.

Recent studies therefore position NDTs as decision-support and pre-validation tools for 6G planning, optimization, troubleshooting, policy verification, and AI/ML lifecycle support~\cite{karamchandani_applicability_2025,calvillo-fernandez_attention_2025,xiao2026nativeintelligence6gllmtrained}. In this role, the NDT acts as an intermediate reasoning layer between live telemetry and control. It allows alternative actions to be explored before affecting the production network, supports the comparison of AI-driven decisions, and improves the robustness of closed-loop management.

This role is especially relevant when network optimization depends on information that is not available through conventional telecom interfaces. Sustainability-oriented operation is a representative example: energy-aware policies can be derived from traffic and power-state information, but carbon-aware policies also require external indicators such as renewable energy availability, grid carbon intensity, and energy forecasts. This extends the NDT beyond domain-internal modelling and motivates architectures capable of combining telecom and non-telecom information sources in a consistent decision process.

\subsection{Classical NDT vs Multi-Domain NDT}
A useful way to position the literature is to distinguish between classical, domain-scoped NDTs and multi-domain NDT approaches. Classical NDTs typically serve as domain-specific capabilities limited to a particular operational segment, such as RAN, transport, core, edge/cloud, or service management. They provide mature mechanisms for modelling, synchronization, analytics, and what-if evaluation within bounded environments. Moreover, domain-specific NDTs can preserve provider privacy by exposing only abstracted models, capabilities, or simulation outcomes, without requiring each domain owner to disclose detailed network configurations or proprietary operational data. However, they usually do not address how decisions spanning several technological or administrative domains should be decomposed, distributed, evaluated, and consolidated.

Multi-domain NDT approaches address this broader scope by coordinating multiple models, simulators, or twin capabilities across heterogeneous environments rather than assuming a single monolithic twin. This direction is aligned with the operational reality of future 6G systems, where end-to-end services depend on the joint behavior of radio, transport, core, cloud-edge, and external domains. The key challenge is therefore not only to construct accurate domain-level twins, but also to define how an end-to-end intent or what-if request is translated into domain-specific analyses and how their outputs are aggregated into an actionable operational decision. In addition, coordination mechanisms must also account for privacy and confidentiality constraints, ensuring that domain providers can participate without exposing detailed network configurations, internal policies, or proprietary operational data.

Table~\ref{tab:classical_vs_multidomain_ndt} summarizes representative works from this perspective. Each entry is characterized by the domains it addresses, covering functional, technological, or administrative scopes across network segments such as \ac{RAN}, transport, core, and edge/cloud. The comparison highlights a clear trend: while scoped \glspl{NDT} provide mature mechanisms for modelling and evaluation within bounded environments, multi-domain approaches extend the architectural scope but still lack a systematic framework for inter-domain coordination, result consolidation, and end-to-end decision support.

\begin{table*}[htp!]
\caption{Comparison of classical and multi-domain NDT literature}
\label{tab:classical_vs_multidomain_ndt}
\centering
\footnotesize
    \begin{tabular}{|p{1.8cm}|p{4cm}|p{5.4cm}|p{5.0cm}|}
    \hline
    \textbf{Class} & \textbf{Representative works} & \textbf{Domains and segments} & \textbf{Positioning w.r.t. this paper} \\
    \hline
    Classical NDTs & 
    Nguyen \emph{et al.} \cite{nguyen_digital_2021}, 
    Sanz Rodrigo \emph{et al.} \cite{sanz_rodrigo_digital_2023}, 
    Nardini and Stea \cite{nardini_enabling_2024},
    Vil\`a \emph{et al.} \cite{vila_design_2023}, 
    Apostolakis \emph{et al.} \cite{apostolakis_digital_2023}, 
    Vilalta \emph{et al.} \cite{vilalta_applying_2023}, 
    Cherini \emph{et al.} \cite{cherini_building_2025}, 
    Polverini \emph{et al.} \cite{polverini_digital_2023} &
    Mobile operator, slice mgmt., simulation/service across RAN, core, and edge/cloud segments, optical transport, and routing/control domains, and IP/BGP segments &
    NDT as support for 5G/B5G operation, synchronization, and simulation-based evaluation; bounded to single operator/system view, lacking inter-domain orchestration. \\
    \hline
    Multi-domain NDTs &
    Wang \emph{et al.} \cite{wang_6g_2024}, 
    Robitzsch \emph{et al.} \cite{robitzsch_standardisation_2024}, 
    Giardina \emph{et al.} \cite{giardina_hierarchical_2023},     
    Faye \emph{et al.} \cite{faye_integrating_2024}, 
    Zaki-Hindi \emph{et al.} \cite{zaki-hindi_reference_2026}, 
    Raza \emph{et al.} \cite{raza_comprehensive_2025} &
    Service, logical, and resource domains across cross-layer 6G architectures and end-to-end service paths &
    Introduce cross-domain coordination across heterogeneous environments; partial definition of inter-twin orchestration and result consolidation. \\
    \hline
    \end{tabular}
\end{table*}

Overall, the literature indicates that current NDT research is progressing from bounded, segment-specific realizations toward broader multi-domain architectures. However, this transition remains incomplete. Existing works do not yet fully specify how end-to-end decisions should be translated into distributed twin queries, how heterogeneous simulation outputs should be consolidated, or how the resulting recommendations should be exposed to orchestration and closed-loop control functions. This gap directly motivates our work, which focuses on the role of the Multi-Domain Manager as the coordination function responsible for interacting with domain-specific twin capabilities and consolidating their outputs for end-to-end decision support.

\section{Background: Multi-Domain Orchestration Architecture for 6G Networks}
\label{sec:background}

Future wireless network environments, especially 6G systems, require a shift from traditional single-domain management to highly federated, multi-domain ecosystems. These networks will integrate diverse access technologies, non-terrestrial components, and complex service chains distributed from deep edge deployments to centralized clouds. To manage the scale and heterogeneity of these cross-domain environments, the UNITY-6G project ~\cite{UNITY6G} proposes a distributed architecture based on the  \ac{SBMA} approach presented in~\cite{3gpp_ts28533}.  

\begin{figure}[htp!]
    \centering
    \includegraphics[width=0.8\linewidth]{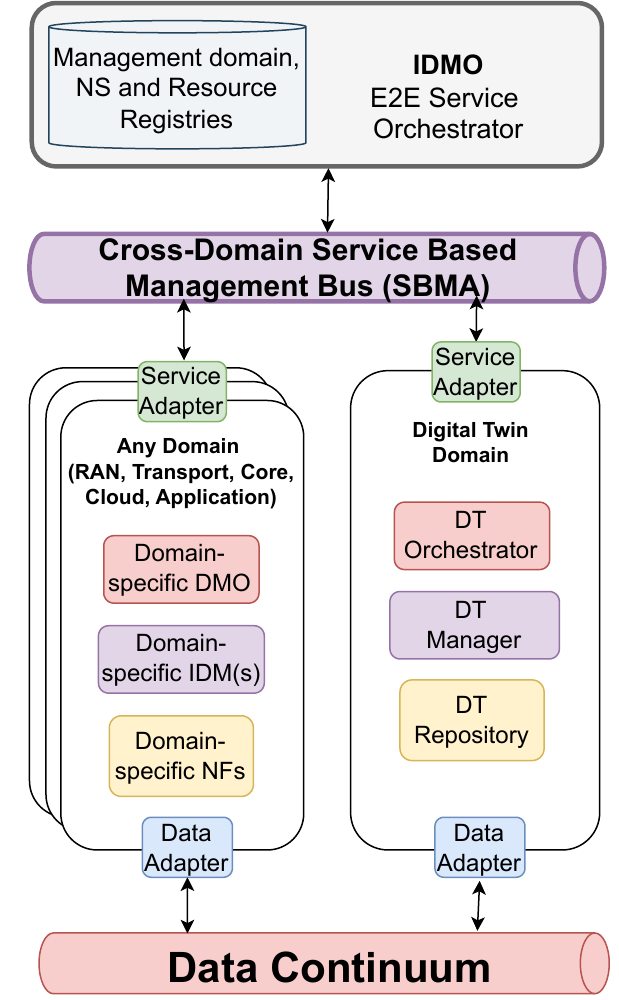}
    \caption{Multi-domain Orchestration architecture for unified 6G networks 
    }
    \label{fig:idmo-control-architecture}
\end{figure}

At the core of this distributed paradigm, a layered control structure decouples global service composition from localized infrastructure execution. As shown in Figure~\ref{fig:idmo-control-architecture}, the \ac{IDMO} entity is at the top of the architecture. The IDMO oversees multiple administrative and technological domains to establish end-to-end (E2E) network services at the highest level of hierarchical control. When an E2E service intent request is received, the \ac{IDMO} is responsible for decomposing the high-level intent into functional components (e.g., RAN, Transport, Mobile Core). It then executes a multi-criteria decision-making process to select the optimal destination domains based on available service capabilities, resource availability, and latency or coverage suitability.

At each selected domain, the \ac{DMO} receives, via the Service Adapter module, the decomposed service component to perform its \ac{LCM}. The \ac{DMO} optimizes intra-domain resources, deploys virtualized network functions (VNFs/CNFs) at selected \glspl{PoP}, monitors domain and service-level KPIs, and reports status to the \ac{IDMO} through interaction with the underlying \glspl{IDM}, which operate the computing resources at the infrastructure level. All communication related to management, orchestration, and policy synchronization between the \ac{IDMO} and domains occurs over the Cross-Domain \ac{SBMA} bus. This vertical approach enables the IDMO to maintain global coordination while delegating local orchestration and infrastructure control to the DMOs and IDMs, respectively, reducing the computational workload of decision making. In later sections, we explain how our \ac{DT} architecture integrates into this broader network architecture and which components replace the general structure.

\section{Proposed Approach}
\label{sec:proposed-approach}

Given the increasing complexity and heterogeneity of future wireless networks \cite{tarneberg20226g}, accurately simulating the entire network continuum is computationally prohibitive. The diversity of user equipment, proprietary hardware, and domain-specific variables makes relying on a single, monolithic simulator impractical for actionable what-if analysis. As a result, the development of holistic \ac{DT} is significantly hindered, since maintaining real-time, end-to-end fidelity across all network domains is extremely difficult to achieve in practice, especially in real time.

Nevertheless, because service delivery spans highly interdependent segments, including the RAN, transport layers, mobile core, and distributed edge-cloud infrastructures, isolated domain-specific testing is insufficient. Optimizing these networks requires accounting for cross-domain effects, such as a RAN configuration shift directly impacting edge computational loads. Therefore, a multi-domain NDT is essential to provide the comprehensive network visibility necessary to optimize interdependencies and guarantee deterministic performance across the entire service path. Accordingly, in this section we describe our proposed architecture, which can be broken down by task, and our query-based workflow that uses this architecture to resolve DT-related queries.

\subsection{DT Architecture}
To overcome the mentioned constraints, the proposed architecture adopts a pragmatic, query-driven approach: rather than maintaining a monolithic continuous simulation, a \ac{DT} Manager receives a what-if query, identifies which available simulators are relevant to that query's domain requirements, invokes them with the appropriate parameters, and consolidates their outputs into a unified result.

The composition of simulators is determined by the query, not pre-configured in advance, which allows the same architectural framework to serve radically different scenario types by selecting different functional subsets. This design is a direct architectural response to the impossibility claim above, because no single function can span all domains. An important detail of this design is that the orchestration layer can natively ingest external, non-telco contextual information (e.g. local renewable energy availability and smart grid telemetry) by treating simulators on equal footing with digital twin modules within the same composition framework.

\begin{figure*}
    \centering
    \includegraphics[width=1\linewidth]{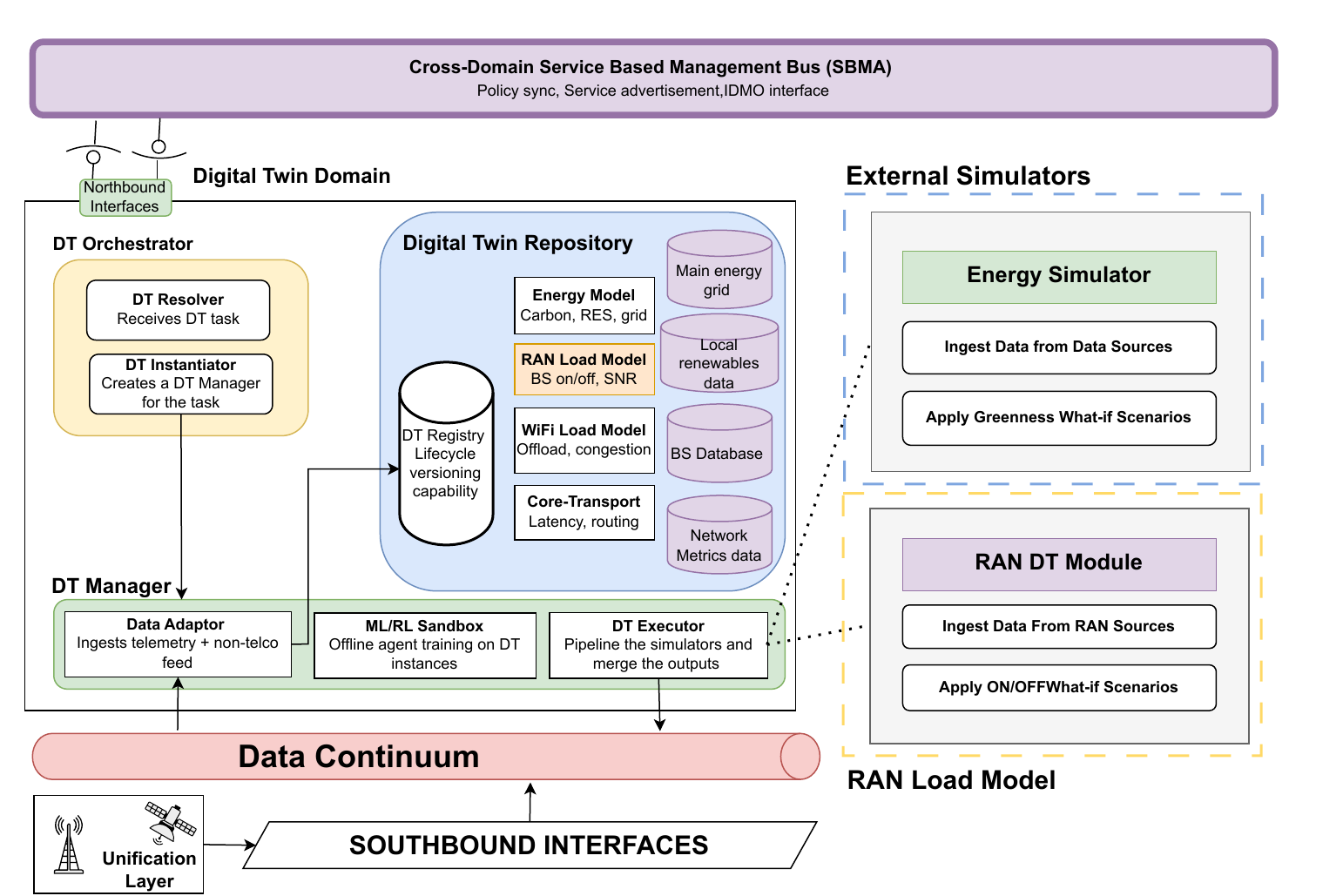}
    \caption{Overall digital architecture and corresponding components interactions. Northbound interfaces represent how the applications interact with the DT whereas Southbound interfaces represent where the twin data comes from~\cite{ITUY3090}.
    }
    \label{fig:dt-architecture}
\end{figure*}

Figure~\ref{fig:dt-architecture} illustrates the internal organisation of the \ac{DT} domain and its interactions with the selected future wireless networks architecture (See Section~\ref{sec:background}) multi-domain architecture. At the top, the Cross-Domain Service-Based Management Bus (SBMA) serves as the northbound communication plane through which the other domains forward what-if tasks and receive the DT's response. A task may be \textit{predictive} (e.g., \textit{what happens if I install 25 solar panels?}), to which the service replies with evaluation results, or \textit{prescriptive} (e.g., \textit{which configuration maximizes greenness under a fixed panel budget?}), to which the DT replies with a recommended configuration. In either case, enforcement authority remains with the requesting entity.

Within the DT domain, the \emph{DT Orchestrator} acts as the control entry point: its \emph{DT Resolver} receives incoming tasks from the SBMA, while the \emph{DT Instantiator} creates a dedicated \emph{DT Manager} instance scoped to the specific query so that concurrent, query-isolated twin instances without contention are instantiated. The DT Manager is the central coordination unit. Its \emph{Data Adaptor} ingests both standard network telemetry (e.g. traffic flow, user association) and non-telco contextual feeds (e.g. real-time renewable generation availability, grid carbon intensity) through the data-continuum. The \emph{ML/RL Sandbox} provides an isolated environment for offline training of reinforcement learning agents on synthesised DT instances, decoupling the model development cycle from live network operation. The \emph{DT Executor} pipelines the external simulators (e.g. \emph{Energy Simulator}) with the DT Modules (e.g. \emph{RAN DT Module}) at query time to enable multi-domain what-if reasoning without requiring a unified simulation model to answer a certain query as explained in further sections (See Section~\ref{subsec:simulator-composition}) and consolidates their respective outputs into a unified, actionable result. Alongside the DT Manager, the \emph{\ac{DT} Repository} maintains versioned virtual models spanning energy, RAN, and transport domains, coordinated through a \emph{DT Registry} that governs lifecycle and versioning across successive what-if evaluations. At the bottom of the architecture, the \emph{Data Continuum} and its Southbound Interfaces expose a unification layer that continuously synchronises physical infrastructure state into the Repository, sustaining the fidelity of the virtual models over time.

\subsection{Network Digital Twin Service Workflow Description}
A central design point is that the \ac{DT} operates in parallel with service establishment rather than as a precondition for it: an end-to-end service is first instantiated with a baseline configuration through the standard cross-domain SBMA procedure sketched in Section~\ref{sec:background}, and the \ac{DT} subsequently derives refined configurations while drawing on telemetry from the already running deployment. This avoids the impasse of having to identify a configuration from a system that is not yet operational and ensures that the \ac{DT} reasons over a measurable, live state.

When an end-to-end intent expresses an analytical requirement (e.g., maximizing renewable utilization under a bounded outage), the inter-domain coordinator decomposes the intent into per-domain components. The RAN, Transport, and mobile Core components are dispatched to their respective appropriate domain, where \glspl{DMO} bring up the corresponding service component, while a separate \ac{DT} sub-request carrying only the analytical intent (target \acp{KPI}, scope, and time horizon) is forwarded through the SBMA to the \ac{DT} domain. Domain selection reuses the capability-based mechanism (See Section~\ref{sec:background}) where the \ac{DT} domain advertises its modelling capabilities (e.g., RAN, Core, Transport), and is selected when those capabilities cover the requested scope. The coordinator does not specify which simulators to run or how to compose them, so the workflow is not bound to a specific inter-domain coordinator implementation and can equivalently be initiated by another domain holding that role.

Composition is therefore a responsibility of the \ac{DT} Orchestrator. The DT Resolver interprets the analytical intent and consults the DT Registry to identify the virtual models and simulators in the DT Repository that match the requested scope; the DT Instantiator then allocates a query-scoped DT Manager. The DT Manager populates the selected models with topology, configuration, and live telemetry obtained from the running service through the Data Continuum, complemented when needed by non-telco feeds such as renewable generation forecasts or grid carbon intensity. The DT Executor invokes the composed simulators and consolidates their outputs. This is also where cross-domain effects are captured: an individual \ac{DMO} is by construction limited to the visibility of its own domain, whereas the \ac{DT} Orchestrator selects simulators that jointly cover the relevant domains and the DT Manager combines their results into a single evaluation.

Candidate configurations are ranked against the requested \acp{KPI} and the best-performing one is returned to the \ac{DT} Orchestrator, which forwards it through the SBMA to the execution \glspl{DMO} that own the affected resources (RAN, Core, Transport, or any combination thereof). Each receiving \ac{DMO} enforces the configuration through its associated \glspl{IDM}. The resulting operational state is reflected back into the Data Continuum and feeds subsequent \ac{DT} evaluations, closing the loop. DT Service Request box illustrates a concrete instance of this workflow.

\subsection{Network Twinning}

Network twinning denotes the process by which selected operational evidence from the deployed system is transformed into an executable representation that can be used by the \ac{DT} environment. In the proposed architecture, this transformation is not limited to copying static inventory data, but relies on exposure mechanisms capable of presenting the real system in a structured and machine-consumable form. For instance, when an operational domain is controlled or observed through \ac{SDN}, an interface such as TerraFlow SDN \cite{etsi_teraflowsdn} can serve as an abstraction layer between heterogeneous network resources and the \ac{DT} environment. Instead of requiring the simulator to interact directly with individual devices, vendor-specific management systems, or low-level configuration procedures, the controller exposes a normalized view of the managed domain through programmable interfaces. This view captures the operational information required for the requested evaluation and presents it in a form that can be interpreted by the twinning function.

The twinning function then maps this exposed operational state into the modelling primitives required by the selected \ac{DT} modules or simulators. In the case of an NS-3-based \cite{ns3_riley_henderson_2010} component, this may involve translating the exposed system view into simulated entities, relationships, parameters, and runtime conditions that correspond to the relevant part of the real network. The same principle applies to other domain models: the objective is not to reproduce the entire deployed system, but to instantiate a scoped executable model whose structure and parameters are grounded in the current or recent operational state. In this sense, TerraFlow SDN and similar exposure functions make twinning possible by converting heterogeneous managed resources into a consistent control-plane representation that can be consumed, parameterized, and updated by the \ac{DT}. The resulting twin is therefore not a passive database replica, but a purpose-specific operational model that supports simulation, analysis, and decision validation before actions are applied to the real system.

\begin{tcolorbox}[float,
  floatplacement=t,breakable,
enhanced,
colback=blue!5!white, colframe=blue!75!black,
title=\textbf{\ac{DT} Service Request},
left=1mm,right=1mm,top=1mm,bottom=1mm,
before skip=6pt, after skip=6pt, label={box:dt-request}, arc=3mm, boxrule=0.8pt]

\textbf{Service Intent:} Maximise renewable energy utilisation across the \ac{RAN} within a fixed solar panel budget.

\textbf{Target KPI:}
\begin{itemize}
    \item Minimise grid energy consumption while keeping network outage below \textbf{5\%}
\end{itemize}

\textbf{Context:}
\begin{itemize}
    \item 105 base stations operating under a cell on/off switching policy
    \item Local solar generation available; grid carbon intensity varies hourly
\end{itemize}

\textbf{DT Task:}
\begin{itemize}
    \item Simulate network load and BS operational states under the active coverage policy
    \item Evaluate solar dispatch scenarios across candidate BS deployments and panel capacities
\end{itemize}

\textbf{Expected Output:}
\begin{itemize}
    \item Optimal set of BSes for solar panel installation maximising grid displacement
    \item Projected grid energy reduction per deployment size
\end{itemize}

\textbf{Action:}
\begin{itemize}
    \item Forward recommended BS configuration via \ac{SBMA} to the RAN domain for execution at \ac{DMO}
\end{itemize}
\end{tcolorbox}

\section{Green network use case}
\label{sec:green-network-uc}

Realizing the sustainability goals of modern telecommunication networks requires a paradigm shift from simply managing operational energy efficiency to actively expanding localized renewable generation. A fundamental problem in 6G service deployment methodologies is their highly siloed nature where they focus almost exclusively on optimizing coverage, capacity, and internal traffic loads. Consequently, they remain "blind" to the external spatial and environmental factors that present physical opportunities to make the infrastructure greener, specifically the untapped solar potential of existing base station (BS) sites and their surrounding topologies.

By strategically deploying solar panels as a supporting instrument for base station energy demand, the network can drastically decrease its dependency on the main energy grid. Not only it is advantageous for the base stations, but it also helps to decrease the stress on the main grid, especially when energy demand peaks during the day. However, planning this integration requires overcoming the divide between telecommunications engineering and environmental modeling. To address this, a dual-simulation approach is essential. First, a dedicated energy simulator is required to evaluate the greenness generation potential considering varying settings. Second, a module representing the network is needed to accurately model the traffic-driven power consumption and other operational constraints of the base stations.

\subsection{Sustainability Aware Simulator}
\label{subsection:sustainability-simulator}
The Energy Simulator is a non-domain-specific simulation component that models the spatiotemporal availability of solar energy and its allocation to a distributed set of demand nodes (e.g. base stations). Unlike a static data lookup, the simulator accepts a parameterized scenario (a candidate deployment or a set of active components, and a time horizon) and evaluates the energy outcomes of that scenario under varying solar energy generation capacities. For the given renewable configurations (i.e, generation capacity, number of panels), the simulator evaluates how it can support the changing energy needs of base stations during the day. Algorithm~\ref{alg:greedy_dispatch} outlines the methodology of this simulator, which operates in two primary stages: (i) candidate node selection and (ii) temporal resource dispatch. 

The first stage (i) identifies target stations that contribute most significantly to system load during critical periods. We evaluate a set of historical system states, each showing different selected base stations to be turned on/off, and their corresponding effect on the energy grid on different times of the day. The goal of this stage is to select candidates that could benefit the most from the dispatched solar energy, therefore it is important to identify the critical system states that demonstrates strong correlation with the hourly patterns of the solar generation. To do this, each set is ranked based on the product of grid stress severity and concurrent solar resource availability. This step isolates periods characterized by high demand and viable resource generation. From the top $N$ critical system states, we identify the active nodes. Each node is assigned a utility score, calculated by multiplying its maximum energy demand during these critical states by its observation frequency (the ratio of states in which the node is active). The nodes are then sorted in descending order based on this score, and we select the top $m$ nodes for resource integration.

The second stage (ii) details the time-step dispatch of the potentially generated solar energy to the selected candidates. The goal of this step is to distribute the hourly generated energy to the different base stations to support the grid during peak energy hours (in other words; peak shaving). For a given solar capacity factor (representing the ratio of how close the actual output can be compared to the generation capacity), we calculate the total available resource for each time interval. Concurrently, the algorithm determines the maximum demand for each candidate node at that specific interval. We then sequentially allocate the available resource to the nodes. Nodes that demand the highest energy are supplied first until the interval's resource generation is fully utilized. We repeat this dispatch process across a range of generation capacities by increasing the number of solar panels to evaluate marginal gains of adding each, and to identify the ideal number of panels that the candidate stations can benefit based on the diminishing returns.

\begin{algorithm}[h!]
\caption{Two-Stage Greedy Heuristic for Resource Dispatch}
\label{alg:greedy_dispatch}
\begin{algorithmic}[1]
\Require Set of all system states $\mathcal{S}$, Set of all nodes $\mathcal{V}$, $N$ (critical states limit), $m$ (candidate limit), $K$ (resource scaling array)
\Ensure Temporal allocation $a_v(t)$ for varying capacities $k \in K$

\Statex \textbf{Stage 1: Candidate Selection}
\For{each state $s \in \mathcal{S}$}
    \State Rank score $R_s \gets \text{stress}_s \times \text{resource\_avail}_s$
\EndFor
\State $\mathcal{S}_{critical} \gets$ top $N$ states sorted by $R_s$ (descending)
\For{each node $v \in \mathcal{V}$}
    \State $f_v \gets \frac{\text{active\_states}_v}{|\mathcal{S}|}$
    \State $U_v \gets 0$
    \For{each $s \in \mathcal{S}_{critical}$ where $v$ is active}
        \State $U_v \gets U_v + (D_{max,v,s} \times f_v)$
    \EndFor
\EndFor
\State $\mathcal{V}_{cand} \gets$ top $m$ nodes sorted by $U_v$ (descending)

\Statex \textbf{Stage 2: Temporal Dispatch}
\For{each capacity scalar $k \in K$}
    \For{each time step $t$}
        \State $R_{total}(t) \gets k \times \text{unit\_capacity} \times \text{resource\_avail}(t)$
        \State Calculate worst-case demand $D_v(t)$ for all $v \in \mathcal{V}_{cand}$
        \State Sort $\mathcal{V}_{cand}$ by $D_v(t)$ in descending order
        \State $R_{remaining} \gets R_{total}(t)$
        \For{each $v \in \mathcal{V}_{cand}$}
            \State $a_v(t) \gets \min(D_v(t), R_{remaining})$
            \State $R_{remaining} \gets R_{remaining} - a_v(t)$
            \If{$R_{remaining} == 0$}
                \State \textbf{break}
            \EndIf
        \EndFor
    \EndFor
\EndFor
\end{algorithmic}
\end{algorithm}
\subsection{RAN DT Module}
The module models a multi-cell 5G network, where base station locations are obtained using real-world data sources, using publicly available base station location databases and BTS search~\cite{btsearch}, to ensure a realistic spatial topology. Each base station is characterized by its transmission parameters, antenna configuration, and maximum power consumption, which are later used in the energy-aware analysis.
The User Equipments (UEs) are modeled using a stochastic mobility process based on a random walk model\, with dynamic spatial traffic distributions over time~\cite{kryszkiewicz2026tandem}. This approach captures temporal and spatial movements in network load, which are critical for evaluating the interaction between traffic demand and renewable energy availability.
At each simulation step, UEs are associated with serving base stations based on signal strength and path loss conditions. The radio resource management follows a round-robin scheduling scheme, allowing the simulator to estimate per-user throughput, SINR, and resource utilization. Based on these metrics, the instantaneous and worst-case load of each base station is computed.
The DART simulator produces time-resolved network states, including base station load, user distribution, and spectral efficiency.

\subsection{Simulator Composition and Query Resolution}
\label{subsec:simulator-composition}
\begin{figure*}[htp!]
    \centering
    \includegraphics[width=.8\linewidth]{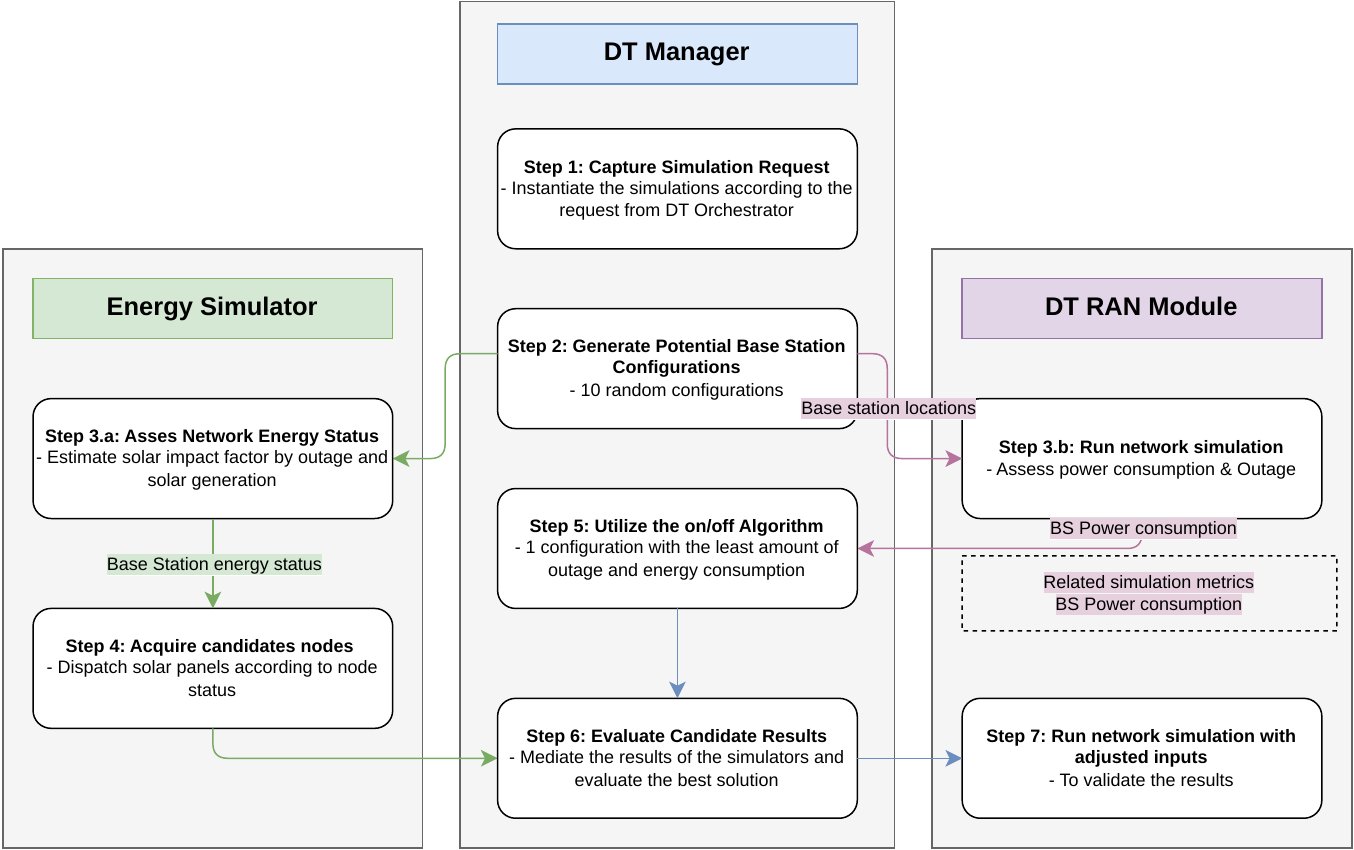}
    \caption{{Process of the interaction between the DT modules, simulators and the DT Manager.}}
    \label{fig:sim-process}
\end{figure*}

Figure~\ref{fig:sim-process} illustrates the interaction between the \ac{DT} Manager, the RAN DT Module and the Energy Simulator for the green networking use case in this study. The process begins when the DT Orchestrator receives  the what-if query from the SBMA and forwards it to the \ac{DT} Manager, which then parses the query and determines the requirements: (i) RAN Simulator for network-layer state generation and (ii) the Energy Simulator for solar allocation as elaborated earlier. In the first phase, O-RAN Simulator is invoked with the base station topology and a stochastic user mobility model. It produces a set of time-resolved network states $\mathcal{S}$, where each state captures per-BS load, active or inactive status, and worst-case power consumption. These states represent the demand-side input that the Energy Simulator cannot determine independently.  In the second phase, the DT Manager passes $\mathcal{S}$, the solar generation profile, and a panel budget parameter $m$ to the simulator. It executes Algorithm~\ref{alg:greedy_dispatch} to identify the $m$ candidate stations and compute temporal solar dispatch, returning the projected grid displacement for each candidate deployment size.

The DT Manager selects the deployment scenario that maximizes grid energy displacement within the panel budget and returns this recommendation as the actionable output to the \ac{IDMO}. The value of this composition is that neither simulator alone can answer the query: O-RAN Simulator cannot evaluate solar allocation, and the Energy Simulator cannot determine which stations are operationally active without the network simulation.

\section{Evaluation}
\label{sec:evaluation}

We apply the generalized framework to localized 5G telecommunications to evaluate the viability of integrating solar photovoltaics (PV) to improve the greenness of the network. The generalized parameters of Algorithm~\ref{alg:greedy_dispatch} are directly mapped to our 5G deployment model. Specifically, the nodes ($\mathcal{V}$) represent clustered 5G network base stations, while the system states ($\mathcal{S}$) include historical network configurations and localized traffic load profiles. We redefine system stress to reflect localized network performance degradation, quantified as the service failure rate (i.e., $1 - \text{service success rate}$, representing unserved traffic or dropped connections). Resource availability remains the hourly solar capacity factor. The product of these metrics identifies timeframes when a base station is under severe traffic load and solar generation is viable. Because nodes experiencing high service failure operate at maximum power draw, targeting them ensures the generated green energy is fully consumed locally by nodes estimated to be under the most duress, maximizing the displacement of grid-derived carbon emissions. Furthermore, the demand ($D_v$) represents the maximum power consumption of a specific base station during a given configuration or hour. Applying this context to the energy heuristic, our model dynamically routes simulated solar power to the most heavily burdened base stations. Iterating through the solar array sizes ($N_{PV}$) determines the marginal utility of expanding physical hardware deployment, ensuring that renewable investments are directed to the nodes where they achieve the greatest reduction in the network's overall carbon footprint.

The Cell On/Off Switching (COOS) xApp (described in~\cite{kryszkiewicz2026tandem}) is located in the Near-RT RIC in the RAN DT Module and is responsible for implementing the logic to turn off certain BSes when necessary. Specifically, each BS $v \in \mathcal{V}$ monitors its traffic load $L_v(t)$ and exchanges load information with its neighboring nodes. A BS is switched off when its load falls below a predefined threshold $\alpha_{\text{off}}$, i.e., $L_v(t) < \alpha_{\text{off}}$, indicating underutilization. Conversely, a previously deactivated BS is reactivated when the average load of its neighboring cells exceeds a threshold $\alpha_{\text{on}}$, i.e., ${L}_{(v)}(t) > \alpha_{\text{on}}$, suggesting potential overload in the surrounding network. Typically, $\alpha_{\text{on}} > \alpha_{\text{off}}$ is assumed to prevent the ping-pong effect. From an energy perspective, this mechanism reduces total network power consumption by eliminating the static power component of underutilized BSes. Traffic consolidation increases the utilization of active nodes, which is more energy-efficient.

\subsection{Assumptions}
\begin{figure}
    \centering
    \includegraphics[width=1\linewidth]{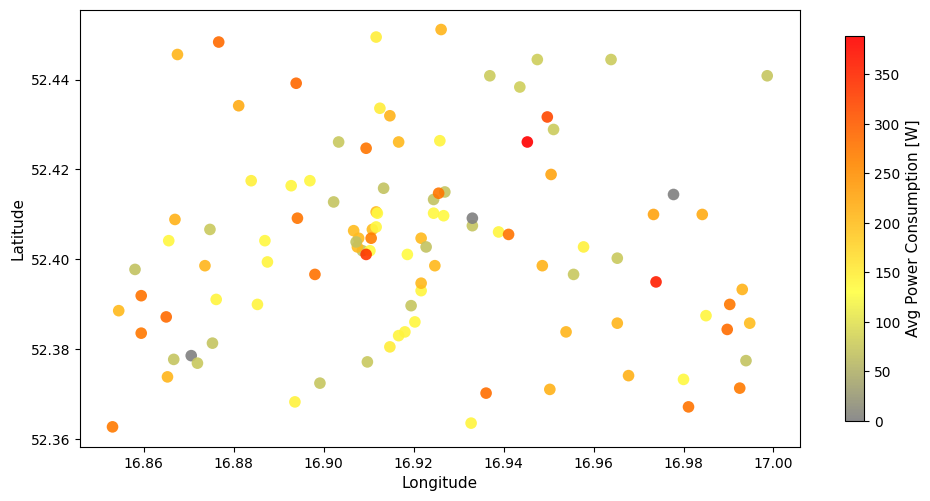}
    \caption{Poznan base station placement and average power consumption during the day.}
    \label{fig:poznan-bs-placement}
\end{figure}

The network topology comprises 105 base stations. Their geographic coordinates, obtained from a real-world database~\cite{btsearch}, are projected onto a local Cartesian reference frame to enable spatial operations, as shown in Fig.~\ref{fig:poznan-bs-placement}. All BSes operate at a carrier frequency of 2100~MHz with a bandwidth of 20~MHz, and a transmit power of 37~dBm, while inter-cell interference is explicitly taken into account. The UEs are randomly distributed over the considered area, with their number varying between 200 and 600 over a 24-hour period to reflect realistic traffic dynamics. Each UE follows a random walk mobility model with an average speed of 5.4~km/h and generates traffic with a constant bitrate demand of 1~Mbps. 

Ten validated BS set configurations (Sets~0-9) represent alternative network operating states with mean outage rates ranging from 8.5\,\% to 15.9\,\% (Table~\ref{tab:outage}) where all the sets have 25 random base stations on.

The coverage optimization algorithm (BS Set ~10) achieves a mean outage of 2.46\,\%, meanwhile the energy optimization algorithm (BS Set ~11) achieves a mean outage of 3.88\,\%, both activating 25 BSes and deactivating the remaining 80. We refer to this as the \textit{Algorithm-25 scenario}. Finally, the candidate selection step of the energy simulation produces a set of BSes similarly comprised of 25 BSes, we refer to the utilization of these BSes in the RAN DT Module as the \textit{Heuristic-25 scenario}.

\begin{table}[h]
\centering
\caption{Mean network outage per BS Set configuration.}
\label{tab:outage}
\begin{tabular}{lc}
\toprule
\textbf{BS Set} & \textbf{Mean Outage (\%)} \\
\midrule
Set~1  &  8.51 \\
Set~4  &  8.89 \\
Set~2  &  9.51 \\
Set~8  & 11.16 \\
Set~0  & 12.19 \\
Set~9  & 12.21 \\
Set~7  & 13.10 \\
Set~6  & 13.94 \\
Set~5  & 15.41 \\
Set~3  & 15.87 \\
Set~10 (Algorithm-25) &  2.46 \\
Set~11 (Heuristic-25) & 3.88 \\
\bottomrule
\end{tabular}
\end{table}

The system integrates real-world grid generation data from the European Network of Transmission System Operators for Electricity (ENTSO-E) with location specific solar generation data obtained from the renewables.ninja platform using the NASA MERRA-2 reanalysis dataset~\cite{pfenninger2016long,staffell2016using}. A single representative measurement point at the Poznan city centre was used, consistent with the homogeneous spatial scale of the study area. The modelled PV output peaked at 0.608\,kW per panel on the simulation date. Solar capacity factors were computed hourly as:
\begin{equation}
    \text{solarCF}(t) = \frac{P_{\text{solar}}(t)}{P_{\text{rated}}} \in [0, 1]
\end{equation}
where $P_{\text{rated}} = 1\,\text{kW}$ per panel (simulator default capacity). The energy dispatch calculations are tested across settings with different number of panels \(k \in [20 , 50]\).

\subsection{Results}

\begin{figure}[h]
\centering
\includegraphics[width=1\linewidth]{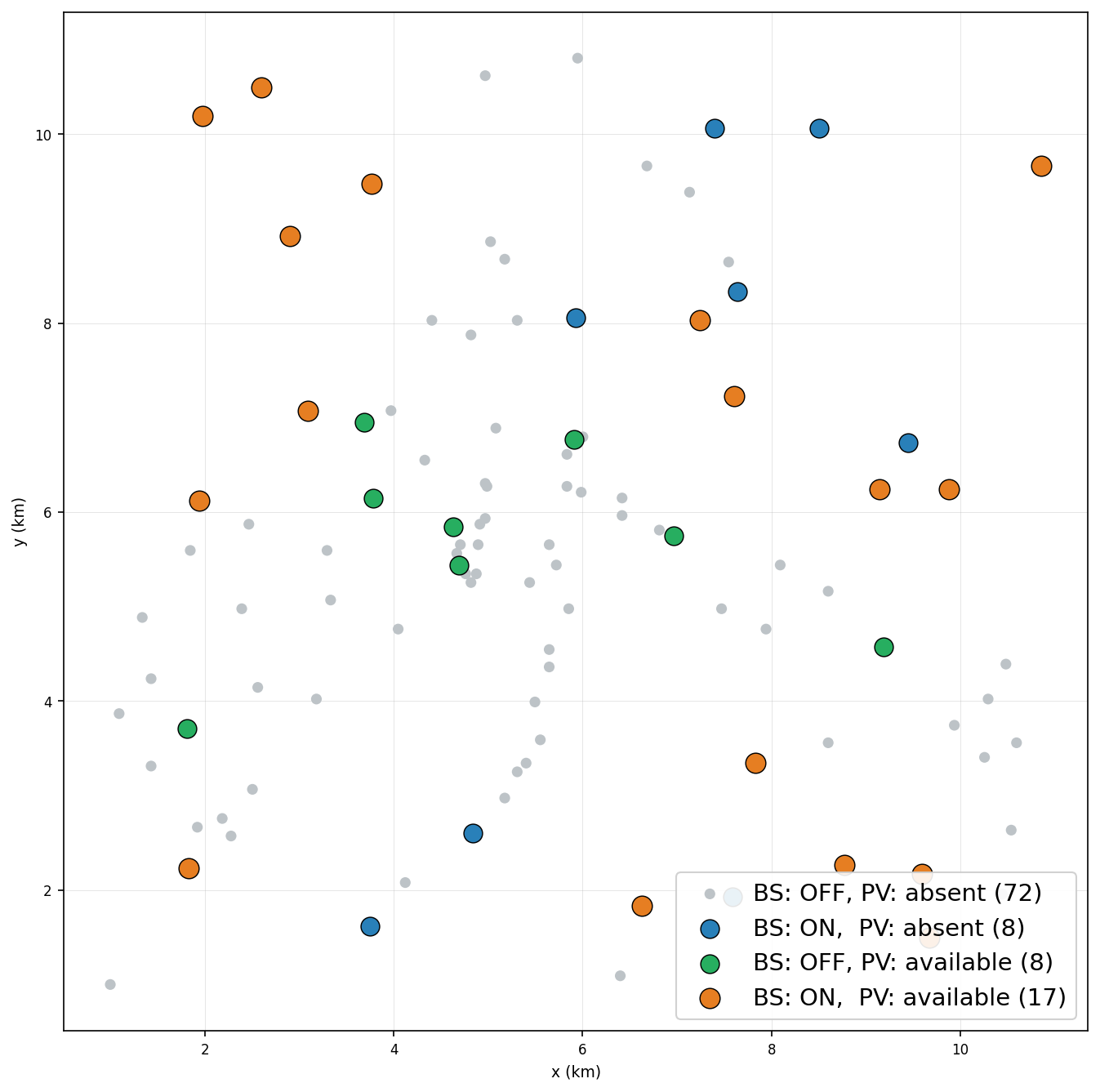}
\caption{BS coverage map. Orange (17 BSes): active in the algorithm and selected by the energy simulator. Green (8 BSes): selected by the energy heuristic but inactive per the O-RAN simulator. Blue (8 BSes): active. Grey (72 BSes): inactive and not selected by either approach.
}
\label{fig:map}
\end{figure}

Figure~\ref{fig:map} shows the geographic location of all the BSes, colour-coded by their joint classification under the two approaches\footnote{All relevant results and data can be found in https://tinyurl.com/yt75257b}.

The orange BSes represent the productive agreement between the two frameworks where 17 stations are both active in the coverage-optimal configuration and ranked as high-priority solar candidates by the energy heuristic. 
The solar panels installed at an orange BS offset real, ongoing grid consumption at a station the network genuinely needs to keep running. The green BSes tell the opposite story where the energy heuristic selected them because they appear active and power-hungry in the random validation configurations, but the coverage algorithm has determined that the network performs better without them. 

The blue BSes represent the these stations are active under the algorithm and consume grid energy throughout the day, yet the energy heuristic did not identify them as solar candidates. They represent the power drawn from the grid that the solar could have offset. The large gray cluster, comprising 72 BSes, is ignored by both approaches so these stations are switched off by the algorithm and ignored by the energy heuristic when making either candidate list.

\begin{figure}
    \centering
    \includegraphics[width=1\linewidth]{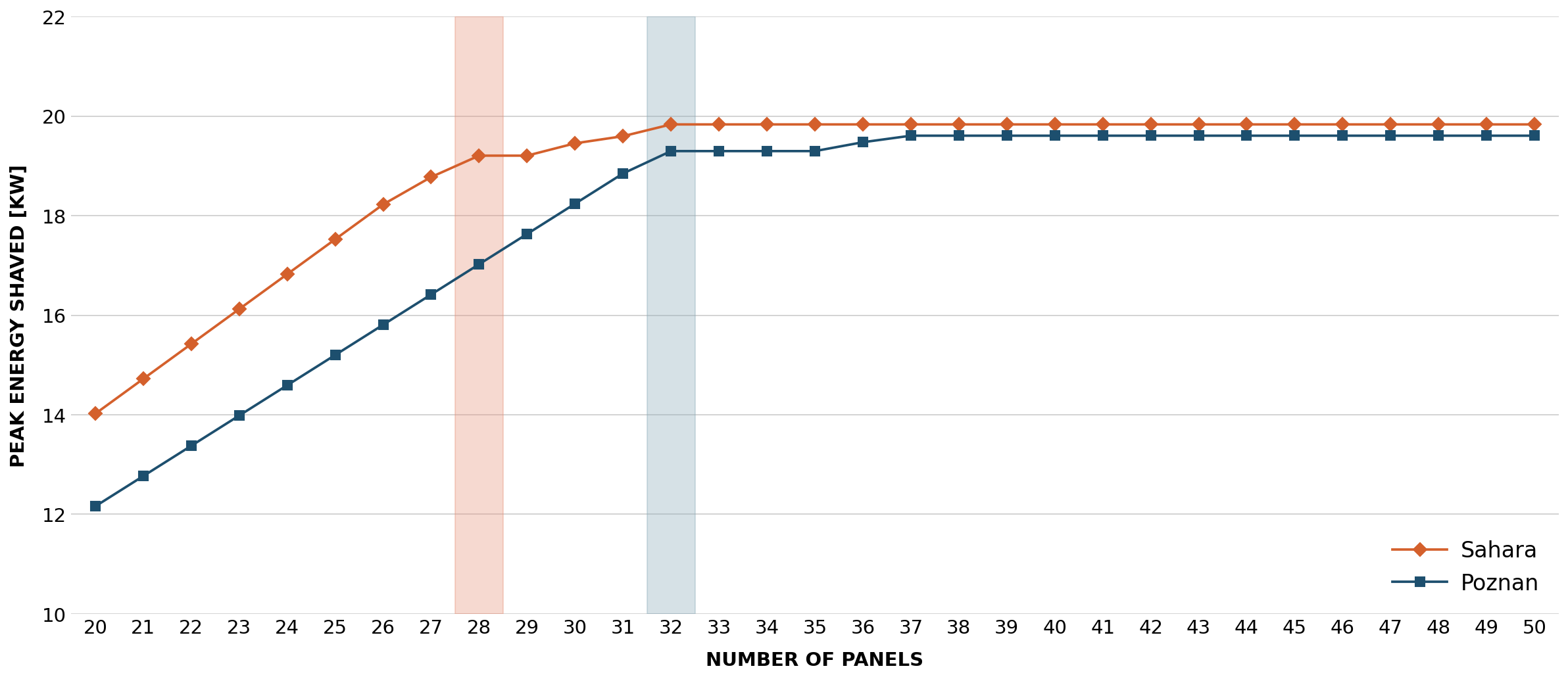}
    \caption{Energy demand peak shaving for Algorithm-25 (Poznan) and same configurations for North Africa (Sahara) use case. Blue rectangle shows the threshold for diminishing benefits of adding a panel for the system in Poznan, whereas orange rectangle shows the same for North Africa use case.}
    \label{fig:peak_shaving}
\end{figure}

Once the energy simulator identifies the base stations to connect to solar energy, the next step is to allocate the expected energy generation to those base stations using varying numbers of solar panels. Since the main goal of this allocation is to support the grid during periods where energy demand is at the peak during the day (i.e., peak shaving), Figure \ref{fig:peak_shaving} shows the total maximum peak energy saved as the number of panels (and consecutively, the total solar energy that can be injected) increases. When considering the total energy consumption needs of the selected base stations operating in the Algorithm-25 scenario, we observe that after 32 solar panels, each additional panel yields diminishing returns.

There are two factors that define the diminishing effect when adding a new panel: i) the total amount of energy that the given set of stations need to consume, and ii) potential solar generation amount depending on the location of the use case. If the same set of base stations with the same configurations had been located in a different region with more solar benefits, we would expect to observe a different number of panels that fits the requirements of the given region. To test this hypothesis, we assume the same network to be operating in the North Africa at the same time period of the original use case. By using our solar energy dispatcher, we allocate the updated amount of solar energy to the base stations with same power consumption and outage values. Similar to the original observation, the results show as the number of panels increase, more peak energy per hour could be relieved from the main grid, until the panels don't contribute to the peak demand periods anymore. For the North Africa use case, we see this critical number of panels is 28, as shown in Figure \ref{fig:peak_shaving}.


The Algorithm-25 scenario establishes the upper bound: among 25 operating base stations, 17 benefit from dispatched solar energy. The stations selected by Algorithm-25 collectively require 458.30 kWh from the grid over the day, of which 32 panels supplying the selected base stations provide 130.4 kWh. Algorithm-25 selects 25 base stations based on network traffic distribution, balancing the need to keep the outage metric low with reducing power consumption. As a result, net grid consumption decreases to 327.9 kWh, corresponding to a 28.5\% reduction.  Alternatively, if all 25 base stations connected to the solar resources operate instead of 17, their total energy consumption over the day becomes 444.76 kWh. With the same solar panel assumptions, the load that can be supplied by solar remains 130.4 kWh, and the net grid consumption decreases by 29.3\%. 

It should be noted that if two distinct sets of operating base stations yield nearly the same total energy consumption, the region's solar potential remains unchanged.  Although the reduction in net grid consumption is a useful metric for assessing the impact of solar potential, Cell On/Off switching decisions are driven by objectives that differ from those of the energy simulator. This complementary simulation helps network operators make long-term decisions to meet sustainability targets while maintaining network performance and reliability.

\section{Conclusions \& Future Work}
\label{sec:conclusions}
This paper presented a \ac{DT}-enabled 6G framework in which NDT capabilities are integrated as a specialized service within a multi-domain orchestration architecture. The proposed approach supports what-if scenario analysis, KPI-driven optimization, and safe validation of candidate actions before deployment in operational networks.
To demonstrate its practical value, a sustainability-oriented use case was developed where network operational data were combined with renewable energy availability to enable carbon-aware base-station operation. Results showed that aligning renewable resource placement with network-aware activation decisions can improve local green-energy utilization while reducing avoidable grid consumption up to 28.5\%, highlighting the benefit of combining telecom and external data sources within future control loops.
Overall, the proposed framework illustrates how NDTs can evolve from passive monitoring tools into active decision-support components for intelligent 6G orchestration. 

A natural extension of this framework is its use in online, fault-driven settings, where choices such as which simulations to run are made on the fly. For instance, the DT Manager could be asked: \textit{``How would network KPIs be affected if the solar panel connected to a base station were to fail?''}. 
This scenario should trigger a coupled simulation in which the energy simulator represents the reduction in generation capacity at that node, while O-RAN Simulator assesses the resulting rise in grid reliance and any loss of coverage under the updated energy budget. 
Such disaster-response scenarios would demonstrate the online, closed-loop potential of the architecture and are a natural next step following the offline planning use case presented here.






\bibliographystyle{IEEEtran}
\bibliography{references}
\balance


\vfill

\end{document}